\documentclass[a4paper, times, 10pt, twocolumn, twoside]{article}
\usepackage{ISARC_arxiv}
\usepackage{lscape}
\usepackage{hologo}
\usepackage{bm}
\usepackage{graphicx}

\usepackage[dvipsnames]{xcolor}

\begin{document}

\linespread{0.5}

\title{Simulation-Based Optimization of High-Performance \\ Wheel Loading}

\author{Koji Aoshima$^{1,2}$, Martin Servin$^{2}$, and Eddie Wadbro$^{2,3}$}

\affiliation{
$^1$Komatsu Ltd., Japan\\
$^2$Umeå University, Sweden\\
$^3$Karlstad University, Sweden
}

\email{
\href{mailto:koji.aoshima@umu.se}{koji.aoshima@umu.se},
\href{mailto:martin.servin@umu.se}{martin.servin@umu.se},
\href{mailto:eddie.wadbro@umu.se}{eddie.wadbro@umu.se},
}

\maketitle
\thispagestyle{fancy}
\pagestyle{fancy}

\begin{abstract}
    Having smart and autonomous earthmoving in mind, we explore high-performance wheel loading in a simulated environment.
    This paper introduces a wheel loader simulator that combines contacting 3D multibody dynamics with a hybrid continuum-particle terrain model,
    supporting realistic digging forces and soil displacements at real-time performance.
    A total of 270,000 simulations are run with different loading actions, pile slopes, and soil to analyze how they affect the loading performance.
    The results suggest that the preferred digging actions should preserve and exploit a steep pile slope.
    High digging speed favors high productivity, while energy-efficient loading requires a lower dig speed.
    \end{abstract}

\begin{keywords}
    Wheel loader; Autonomous; Simulation-Based Optimization; Multibody and soil dynamics
    \end{keywords}

\section{Introduction}
\label{sec:introduction}
    Smart and autonomous earthmoving equipment may significantly improve energy efficiency, productivity, and safety
    at construction sites and mines. If the planning and control system can be made well-informed about the
    physics of earthmoving operations and the current state of the environment, then it can predict the outcome
    of an action and select near-optimal action sequences that are well-coordinated with other systems at the
    site. This motivates us to explore which wheel loading actions that maximize the performance over a task.
    The loading task is typically operated as a repeating cycle of sequential actions: heading into a pile,
    scooping, breaking out of the pile, carrying the soil, and dumping it to fill bodies of dump trucks
    \cite{Hemami2009}.
    \begin{figure}
        \centering
        \includegraphics[width=0.30\textwidth]{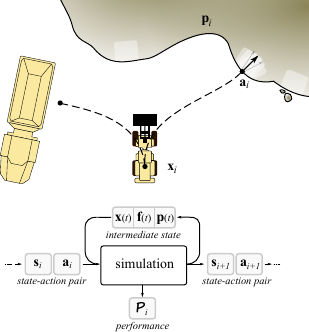}
        \caption{The typical loading cycle of filling a bucket and dumping the material in a receiver can be
        viewed as selecting a control action that transforms the system from one state to another with some
        performance that depends on the intermediate states.}
        \label{fig:loading_actions}
    \end{figure}
    %
    Despite the repetitive task, adjusting the loading actions to the environment for consistent high-performance loading is challenging.
    The difficulty lies in the complex dynamics and variability in the wheel loader-soil interaction.
    It is impractical to address the challenge through physical experiments.
    Systematic and repeatable experiments appear possible only in simulated environments,
    but computationally efficient models for wheel loading with realistic soil dynamics have become available only recently. 

    In this paper, we introduce a method for exploring the sequential loading actions for maximum performance in a simulated environment.
    We view it as an optimization problem and analyze how the loading actions are expected to perform according to the situations.
    A simulator is developed, which combines contacting 3D multibody dynamics with a hybrid continuum-particle terrain model
    that supports realistic digging forces and soil displacements at real-time performance \cite{Servin2021}.
    To show the ability of the simulator, we demonstrate how the loading performance depends on the pile and the loading action.
    In total 270,000 simulations are conducted with different action parameters
    for loading, pile slopes, and type of material.
    The performance is analyzed statistically, and some characteristic actions are compared in detail.

    \subsection{Related work}
        Some researchers have previously investigated strategies for maximizing the performance of wheel loading
        in simulated environments. 
        They especially focus on the bucket filling action, which is the most predominant action 
        for fuel efficiency and productivity in the loading cycle \cite{Shimizu2019}.
        Singh and Cannon \cite{Singh1998} presented a planning algorithm to search for the best dig region and executed the
        planner by using a machine and pile model in 2D. Sarata et al. \cite{Sarata2005} proposed a method
        for dig planning, taking the 3D pile shape into account and avoiding undesirable stress by unbalanced
        bucket filling.  Filla et al. \cite{Filla2014,Filla2017} investigated optimal bucket filling
        trajectories by using simulation based on the discrete element method (DEM),
        concluding that the ``slicing cheese'' motion, advocated by machine instructors, is indeed a good strategy.
        They also noted the need for fully dynamic simulation models, with the possibility to adapt the control to the
        changing environment as high-performing operators do, for developing optimal bucket filling strategies.
        The previous studies have been limited to either kinematic bucket trajectories or quasi-static
        soil models, often based on fundamental earthmoving equation (FEE) in combination with a cellular automata.
        This has several drawbacks. A kinematic bucket trajectory may not be realizable when the full dynamics
        and actuator force limits are taken into account.  The FEE force resistance does not support situations when
        the vehicle or bucket is accelerating.  The correct loading time and power consumption require a dynamic model.
        Furthermore, quasi-static soil models do not capture the actual soil displacement, around and into the bucket.
        This affects both the bucket fill ratio and soil spillage on the ground, which may significantly penalize the
        performance over time.
        For simulation results to be transferable to the control and planning on real sites,
        the dynamics of both the machine and the soil must be represented in the model.
        That is computationally very challenging, and few studies have been performed.
        Lindmark and Servin \cite{Lindmark2018} explored a control strategy to maximize the loading performance
        using simulation based on 3D nonsmooth multibody dynamics.
        However, it was conducted for a load-haul-dump machine loading fragmented rock.
        Wheel loaders require a different strategy because of a lower breakout force than LHD:s \cite{Dadhich2019} and
        the high-performing loading depends on the pile property \cite{Bradley1998}.

    \subsection{Problem statement}
        During a loading cycle, the wheel loader typically performs a sequence of loading actions as illustrated in Figure~\ref{fig:loading_actions}.
        We assume that the wheel loader has a task planner and a controller for recurring \emph{actions} $(\bm{a}_i)^N_{i=1}$ during $N$ sequential loading cycles.
        A loading action $\bm{a}_i$ starts with a machine state $\bm{x}(t_i) \in \mathbb{R}^{N_\text{l}}$ and a pile state $\bm{p}(t_i) \in \mathbb{R}^{N_\text{p}}$,
         which can be represented as a system \emph{state} $\bm{s}_i = [\bm{x}_i , \bm{p}_i] \in \mathbb{R}^{N_\text{s}}$.
        Then, $\bm{a}_i$ leads to some trajectory $\bm{x}(t)$ and applied force $\bm{f}(t)$, changing the pile state $\bm{p}(t)$ over a time interval $t \in [t_i, t_i + t_\text{load}]$.
        The system ends up with a new state $\bm{s}_{i+1} = [\bm{x}_{i+1} , \bm{p}_{i+1}]$, where
        $\bm{p}_{i+1}$ may include material spilled in the working area.
        The loading cycle can be attributed to a \emph{performance} $\bm{\mathcal{P}}_{i} \in \mathbb{R}^{N_\mathcal{P}}$.
        The performance depends on a state-action pair, that is, $\bm{\mathcal{P}} (\bm{s}_{i}, \bm{a}_i)$.
        Examples of performance measures are the energy efficieny $\mathcal{P}^\text{e} = m_\text{load} / W$, productivity 
        $\mathcal{P}^\text{p} = m_\text{load} / t_\text{load}$, and bucket fill ratio $\mathcal{P}^\text{b} = V / V_\text{bucket}$, where
        $m_\text{load}$ is the resulting mass of the load in the bucket, $W(\bm{x},\bm{f},t_\text{load})$ is the accumulated work excerted by the 
        actuators over the loading cycle of time duration $t_\text{load}$, $V$ is the volume of the load in the bucket that has 
        volume capacity $V_\text{bucket}$. 
        The simulation process is illustrated in Figure~\ref{fig:loading_actions}.
        %
        %
        What is an optimal action sequence $(\bm{a}_i)^N_{i=1}$ may depend on the initial pile state, $\bm{p}_{1}$, and the number of loading cycles, $N$.
        For example, an optimal action sequence for $N \gtrsim 1$ may transform the pile into a poor state.
        That would prohibit continued loading with good performance unless the pile is restored by additional actions optimized for improving the quality.
        On the other hand, an optimal action sequence for over $N \to \infty$ (terminated when the pile is empty) might start with a loading performance
        significantly worse than what is optimal for single loadings but can maintain a good average performance.
        From the perspective above,
        optimization of $N$ sequential loadings correspond to finding $(\bm{a}_i)^N_{i=1}$ that satisfy
        \begin{equation}\label{eq:optimization_problem}
            \max_{(\bm{a}_i)^N_{i=1}} \sum_{i=1}^N \bm{w}^\mathsf{T}\bm{\mathcal{P}}_{i}
        \end{equation}
        where $\bm{w} \in \mathbb{R}^{N_\mathcal{P}}$ are weight factors for the different performance measures $\bm{\mathcal{P}}$.
        Note that optimization over sequential loadings is not carried out in the present paper but will be pursued in future work.   
        %
        %



\section{Simulator}
\label{sec:simulator}

    A simulator is created using the physics engine AGX Dynamics \cite{AGX2021},
    which supports real-time simulation of multibody systems with nonsmooth contact dynamics, driveline, and deformable terrain.

    \subsection{Terrain model}
    The terrain is simulated using a multiscale hybrid model presented by Servin et al. \cite{Servin2021}.
    Resting soil is modeled as a solid, discretized in a regular 3D grid and a corresponding 2D  height map for the free surface that mediates contacts with earthmoving equipment.
    The bulk mechanical properties of the soil are parametrized by the mass density, internal friction, cohesion, and dilatancy at the bank state.
    When earthmoving equipment comes in contact with the terrain, a zone of active soil is predicted and resolved with particles of variable size and mass.
    DEM is used for particle dynamics with frictional-cohesive contact parameters and a specific mass density that matches the bulk parameters of the soil.
    The earthmoving equipment experiences the active soil as a low-dimensional body with the aggregated shape and inertia of the particles and with frictional-cohesive contacts at its interfaces.
    This can be seen as an extension of the FEE to dynamic conditions, including also bucket penetration resistance and contacts between the bucket's exterior and the surrounding soil.
    The multiscale model allows for combined use of a direct solver for the vehicle dynamics at high precision, an iterative solver for scalable particle dynamics at lower error tolerance, and strong coupling between the vehicle and terrain dynamics that resist soil failure when the stresses do not meet the Mohr-Coulomb criteria.

    \subsection{Wheel loader model}
        \begin{figure}
            \centering
            \includegraphics[width=0.30\textwidth, trim = 0 0 0 0, clip]{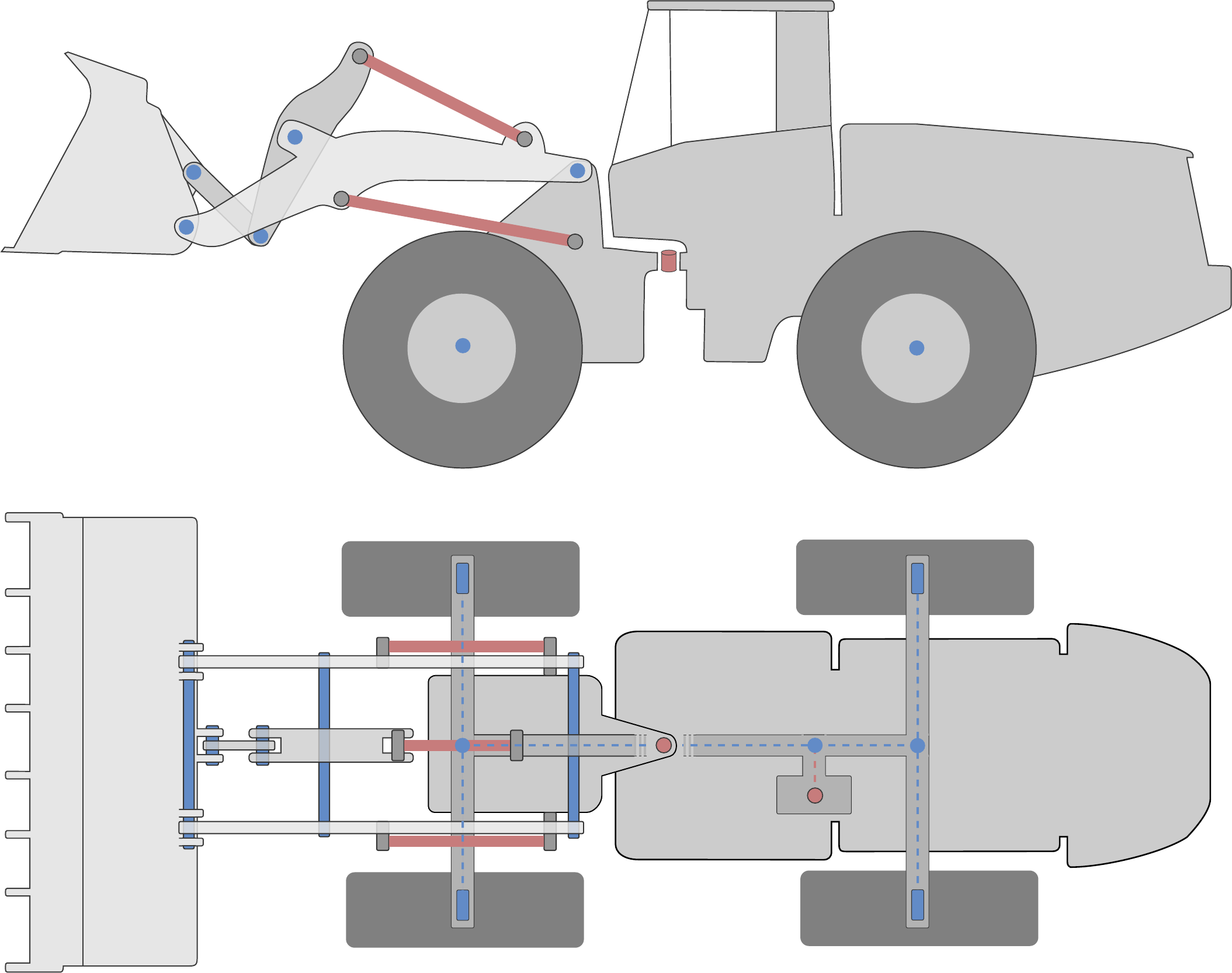}
            \caption{Overview of the wheel loader model with passive joints in blue and actuated joints in red.}
            \label{fig:machine_model}
        \end{figure}
        The wheel loader is modeled as a rigid multibody system consisting of a front and rear frame, connected by a revolute joint for articulated steering,
        four wheels, and a parallel Z-bar linkage system \cite{Oba2013} for controlling the bucket relative to the front frame.
        Bucket filling is the combined effect of thrusting the vehicle into a pile, by applying torque on the wheels, while raising and tilting the bucket.
        The driveline model consists of a revolute motor transmitting rotational power to the front and rear wheel pairs via a main, front, and rear shaft, coupled with differentials.
        Each wheel is connected to the frame with a revolute joint and consists of a tire-hub pair, with finite elasticity with respect to radial, lateral, bending, and torsional displacements.
        The parallel Z-bar linkage system is modeled using 11 revolute and three prismatic joints, with linear motors that represent the hydraulic cylinders for raising and tilting the bucket.
        In total, the assembled model consists of 27 rigid bodies and 23 kinematic constraints, whereof 18 are passive joints and five are actuated.
        The total operating weight is $15.59$ tons, wheelbase $3.030$ m, and the bucket has a volume capacity of $3.0$ $\text{m}^3$.
        Overall, the model roughly corresponds to a Komatsu WA320-7 \cite{KomatsuWA320}.
        The actuators are controlled by specifying a momentaneous joint target speed and force limits, which are listed in Table.~\ref{tab:key_properties}.
        The force limits are chosen to reflect that limited power can be drawn from the engine, which in reality supplies both the hydraulic cylinders and the driveline.
        In addition to the internal constraints, tire-terrain intersections give rise to frictional contact constraints, and bucket-terrain intersections cause soil failure and digging resistance.
        The wheel-terrain friction coefficient is set to $0.8$.
        For clarity of the terrain dynamics, only the wheel and bucket are given visual attributes.
        \begin{table}
            \small
            \caption{Actuators}
            \label{tab:key_properties}
            \centering
            \begin{tabular}{l l l l} \hline
                name    & type      & speed range                 & force limit \\ \hline
                drive   & revolute  & $[0,11]~\text{km/h}$        & $85~\text{kNm}$ \\
                steer   & revolute  & $[-0.1,0.1]~\text{rad/s}$   & $100~\text{kNm}$ \\
                lift    & linear    & $[-0.2,0.11]~\text{m/s}$    & $395~\text{kN}$ \\
                tilt  & linear    & $[-0.2,0.1]~\text{m/s}$     & $530~\text{kN}$ \\ \hline
            \end{tabular}
        \end{table}

    \subsection{Comparison with field measurements}
\begin{figure}
            \centering
            \includegraphics[width=0.45\textwidth, trim = 0 0 0 0, clip]{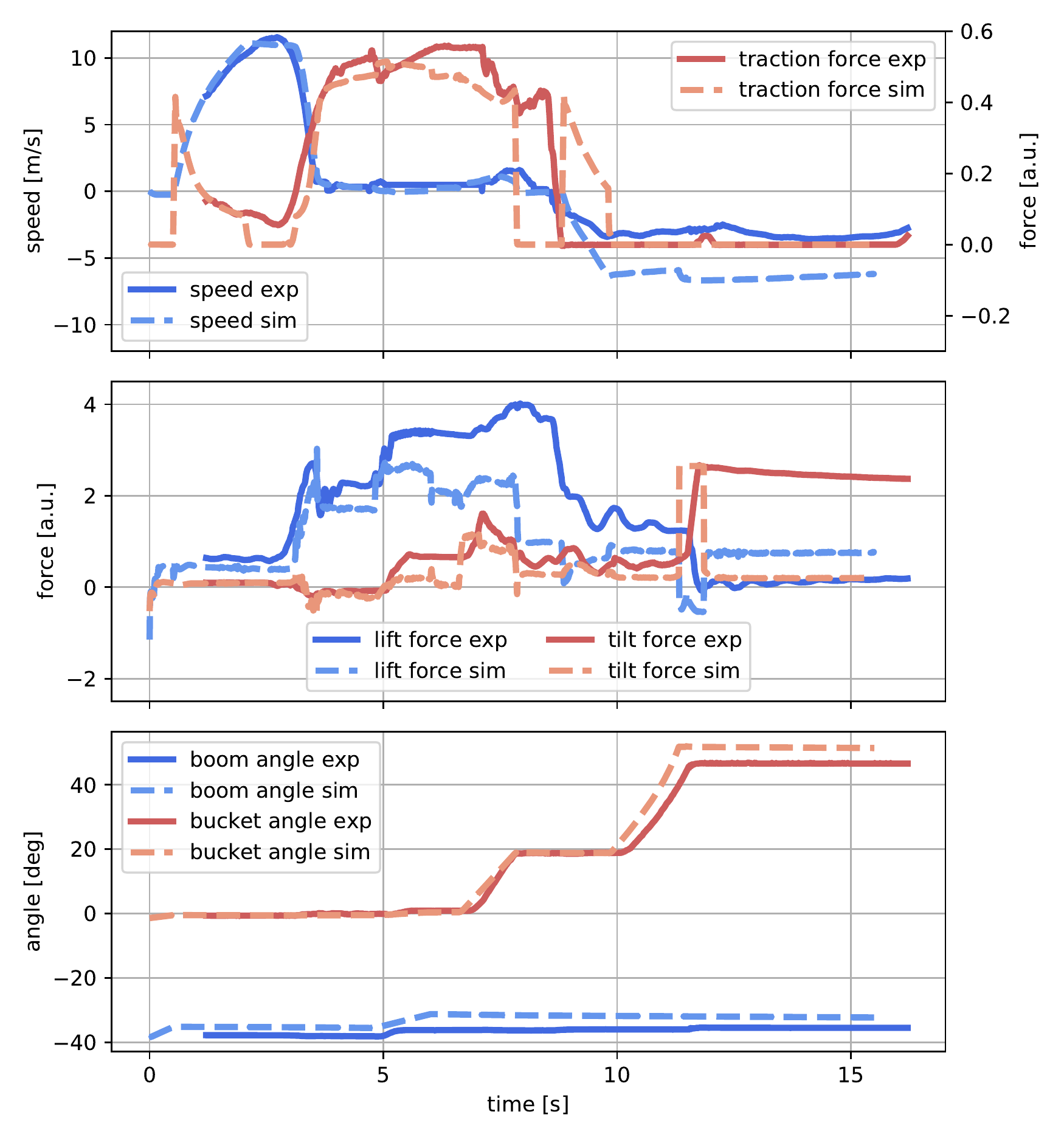}
            \caption{Comparison of the wheel loader speed and forces during a real and simulated loading cycle.}
            \label{fig:comparison_forces}
        \end{figure}

\begin{figure}
            \centering
            \includegraphics[width=0.45\textwidth, trim = 0 0 0 0, clip]{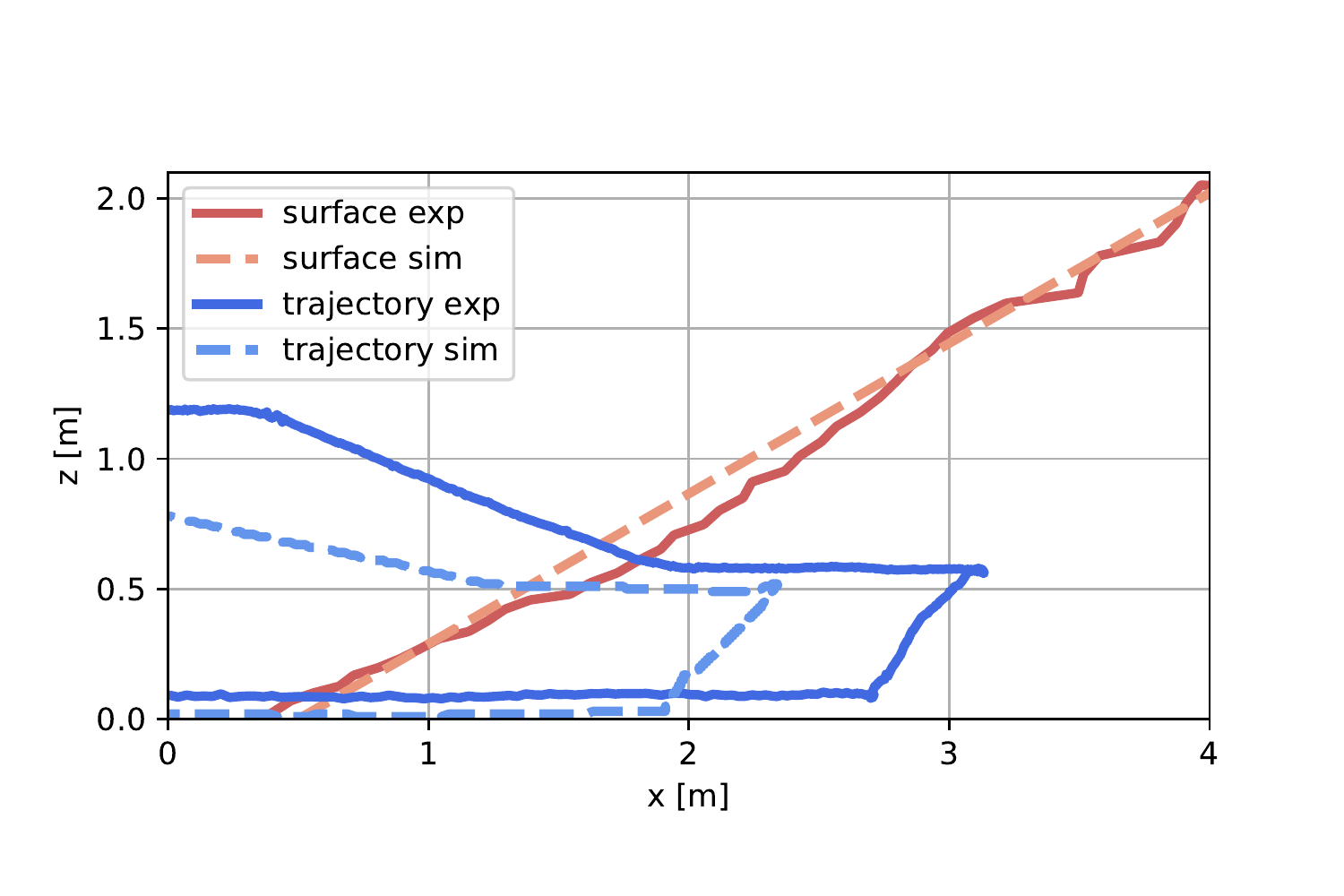}
            \caption{Comparison of the bucket tip trajectory during a real and simulated loading cycle.
            The pile surface is also indicated.}
            \label{fig:comparison_trajectory}
        \end{figure}

    To verify that the simulator represents the real-world counterpart, we compare
    the trajectories and forces of the real and simulated a wheel loader conducting a loading cycle.
    \begin{figure*}
        \centering
        \begin{tabular}{ccc}
          \includegraphics[trim=40 0 0 100, clip, width=45mm]{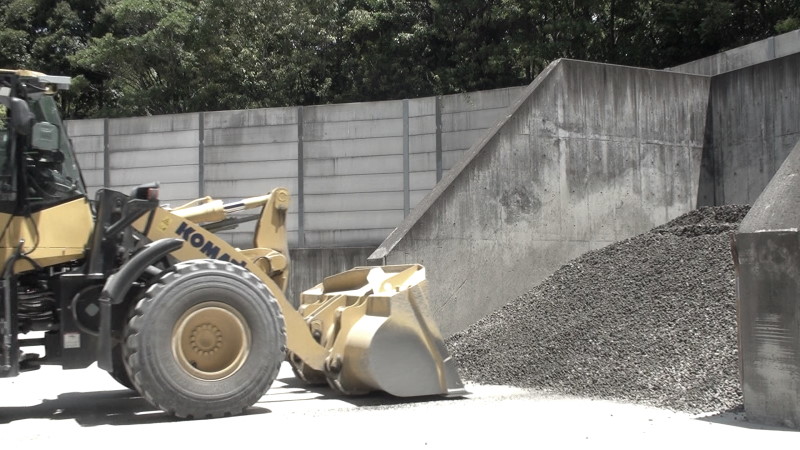} &   \includegraphics[trim=40 0 0 100, clip, width=45mm]{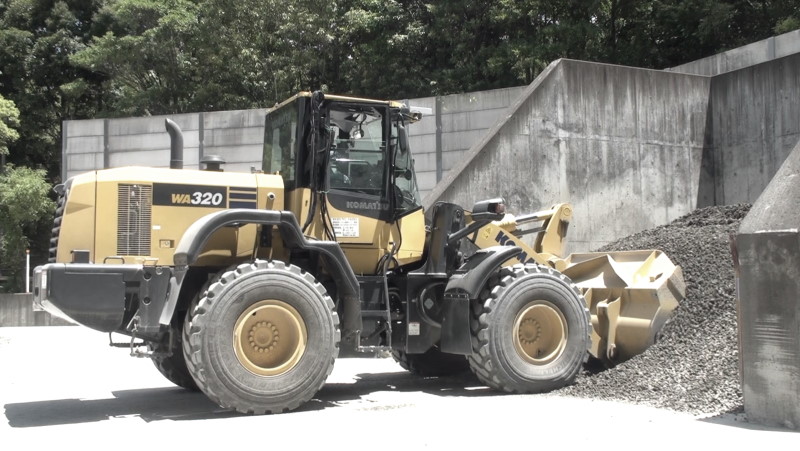} &
          \includegraphics[trim=40 0 0 100, clip, width=45mm]{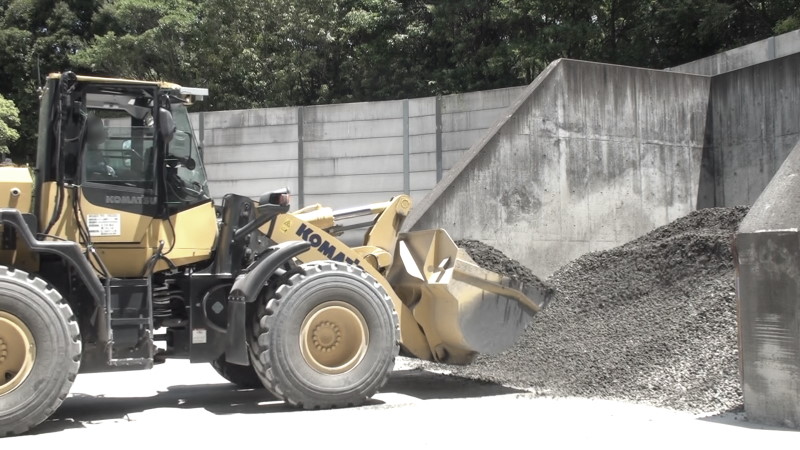} \\ 
         \includegraphics[trim=90 0 0 130, clip, width=45mm]{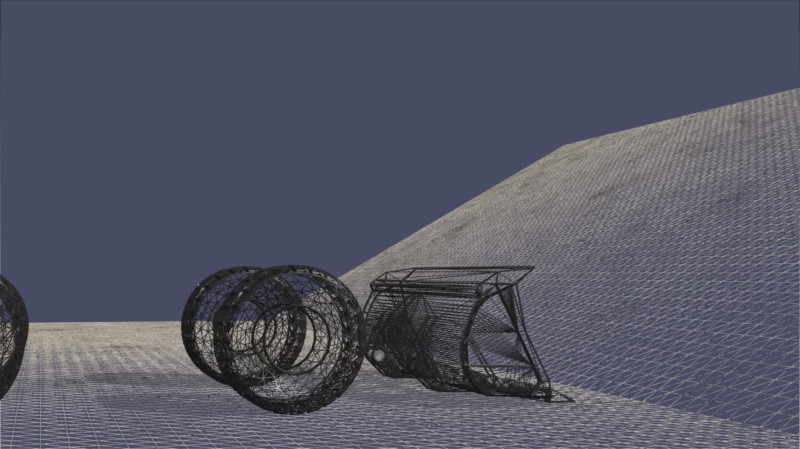} &   \includegraphics[trim=90 0 0 130, clip, width=45mm]{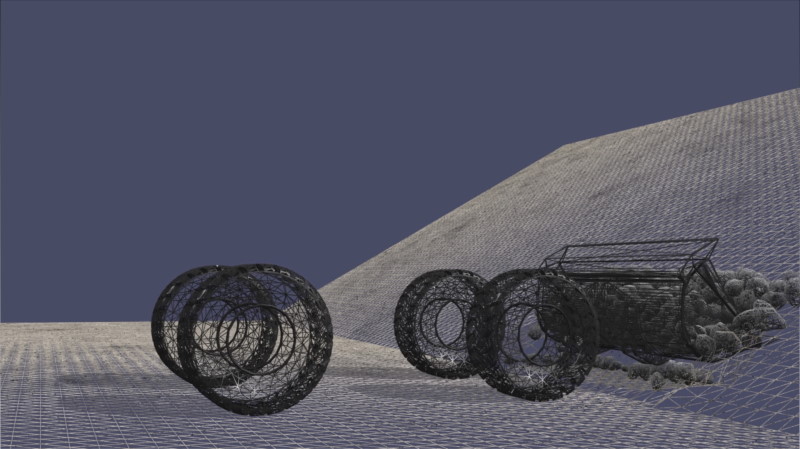} &
         \includegraphics[trim=90 0 0 130, clip, width=45mm]{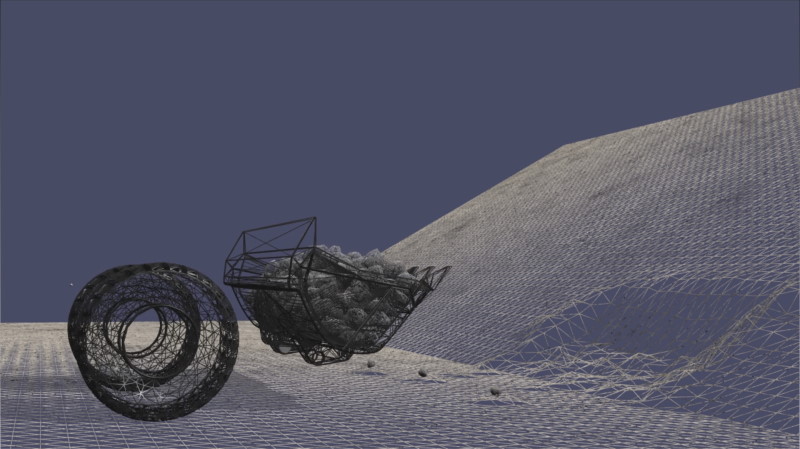} \\ 
        \end{tabular}
        \caption{Image sequence from field experiment and simulation.}
    \end{figure*}
    Data was recorded from a manually operated wheel loader and control parameters in the simulation
    were selected to reproduce a similar, but not identical, loading cycle.
    The vehicle speed, traction force, and lift and tilt cylinder forces are shown in Figure~\ref{fig:comparison_forces},
    and the bucket tip trajectory is in Figure~\ref{fig:comparison_trajectory}. The forces are
    normalized with a characteristic force for the wheel loader.
    The boom and bucket angles relative to the chassis pitch.
    The data in Figure~\ref{fig:comparison_forces} confirms that the model is representative.  The
    forces and trajectories do differ, e.g., the swept area of the real bucket trajectory is about 60\%
    larger than the simulated one.  This agrees with the observed difference in loaded mass and the lift force
    after breakout.  Presumably, the discrepancy is due to differences in pile shape and soil properties, which
    have not been calibrated.

    \section{Simulations}
    \label{sec:simulations}
    The purpose of the simulator is to support the development of high-performance wheel loading from
    different pile states.  Therefore, we run a large set of simulations with different action parameters 
    and analyze their performances in relation to the pile state. 
    Piles with four different slopes are studied: $10^\circ$, $20^\circ$, $30^\circ$ and $40^\circ$.
    The considered soils include gravel, sand, and dirt. Following the terrain library in AGX Dynamics, they have an angle of
    internal friction $44^\circ$, $39^\circ$, and $40^\circ$, and dilatancy $11^\circ$, $9^\circ$, 
    and $13^\circ$, respectively. Gravel and sand are cohesionless,
    while dirt has cohesion $2.1$ kPa. The soils are assigned the same bulk density, $1400$ kg /m$^3$.
    For simplicity, the loading scenario is restricted to entering the pile head-on, with the bucket
    lowered to the ground and reversing straight out after completing the loading. The loading task
    is controlled using eight action parameters, $\alpha_i$, listed in Table 3, and discretized to $45,000$
    parameter combinations per pile.  In total, $270,000$ simulations were run.  Loadings on sand and dirt were 
    carried out only for piles with $30^\circ$ slope.

    The loading is controlled with a simplified version of admittance control \cite{Dobson2017}. The approach
    drive speed is $v_\text{d} = \alpha_1 v^\text{max}_\text{d}$, and the target speed during the
    penetration phase is $v_\text{d} = \alpha_2 v^\text{max}_\text{d}$. When the digging force exceeds the
    set threshold values, $F_{\text{lift}}^{\text{dig}} = \alpha_3 F^\text{dig}_0$ and
    $F_{\text{tilt}}^{\text{dig}} = \alpha_4 F^\text{dig}_0$, the lift and tilt actuators start running with their
    respective target speeds $v_\text{lift} = \alpha_5  v^{max}_\text{lift}$ and
    $v_\text{tilt} = \alpha_6  v^{max}_\text{tilt}$.  This continues until the boom and bucket angles
    reach their target, $\theta_\text{boom} = \alpha_7$ and $\theta_\text{bucket} = \alpha_8$, but
    is aborted in case of breakout or stalling.
    The brake is then applied for one second, letting the
    activated soil come to rest.  After that, the vehicle is driven in reverse at the target speed
    $v_\text{d} = - 0.6 v^\text{max}_\text{d}$ and with tilt target speed
    $v_\text{tilt} = 0.6 v^{max}_\text{tilt}$ until the end position $\theta_\text{bucket} = 50^\circ$
    is reached. After the breakout, the lift target speed is $ v_\text{lift} = 0.6 v^{max}_\text{lift}$
    until the end position $\theta_\text{boom} = -10^\circ$ is reached. The simulation is ended when
    the vehicle has reversed 5 m from the entry point.  The following control constants are used:
    $v^\text{max}_\text{d} = 11.0$ km/h, 
    $v^{max}_\text{lift} = 0.11$ m/s,
    $v^{max}_\text{tilt} = 0.10$ m/s, and $F^\text{dig}_0 = 100$ kN.

    The simulations are run with $10$ ms time-step, with the terrain discretized spatially to $0.2$ m resolution,
    on a high-performance computer with Intel Xeon E5-2690v4 processors, each node enabling up to 28 simulations
    in parallel and roughly $10^4$ loading cycles per CPU hour. 
    During each simulation, the position, velocity, and force are registered over time for selected bodies, joints, and actuators.
    From this, the loaded mass $m_\text{load}$, dig time $t_\text{load}$, and energy consumption $W$ are computed for each loading cycle,
    as well as the relative load spillage $s_\text{load} = V_\text{spill}/V_\text{bucket}$ of material on the ground, and the resulting pile shape.
    The energy efficiency and productivity performance measures, $\mathcal{P}_\text{e}$ and $\mathcal{P}_\text{p}$, 
    are computed for each loading also.

        %
        \begin{table}
            \small
            \centering
            \caption{Action parameters}
            \label{tab:action_parameters}
            \begin{tabular}{l l l} \hline
                            & control           & values \\  \hline
                $\alpha_1$  & approach speed    & $[0.4, 0.6, 0.8]$  \\
                $\alpha_2$  & penetration speed & $[0.2, 0.4, 0.6]$  \\
                $\alpha_3$  & lift-trigging dig force & $[0.0, 0.3, 0.6, 0.9, 1.2]$ \\
                $\alpha_4$  & tilt-trigging dig force & $[0.0, 0.3, 0.6, 0.9, 1.2]$  \\
                $\alpha_5$  & lift speed & $[0.2, 0.4, 0.6, 0.8, 1.0]$ \\
                $\alpha_6$  & tilt speed & $[0.2, 0.4, 0.6, 0.8, 1.0]$ \\
                $\alpha_7$  & lift angle & $[-40^\circ, -30^\circ, -20^\circ, -10^\circ]$ \\
                $\alpha_8$  & tilt angle & $[30^\circ, 45^\circ]$ \\ \hline
            \end{tabular}
        \end{table}

\section{Result}
\label{sec:result}
\begin{figure*} [t]
        \centering
            \includegraphics[trim=8 11 15 17, clip, width=40mm]{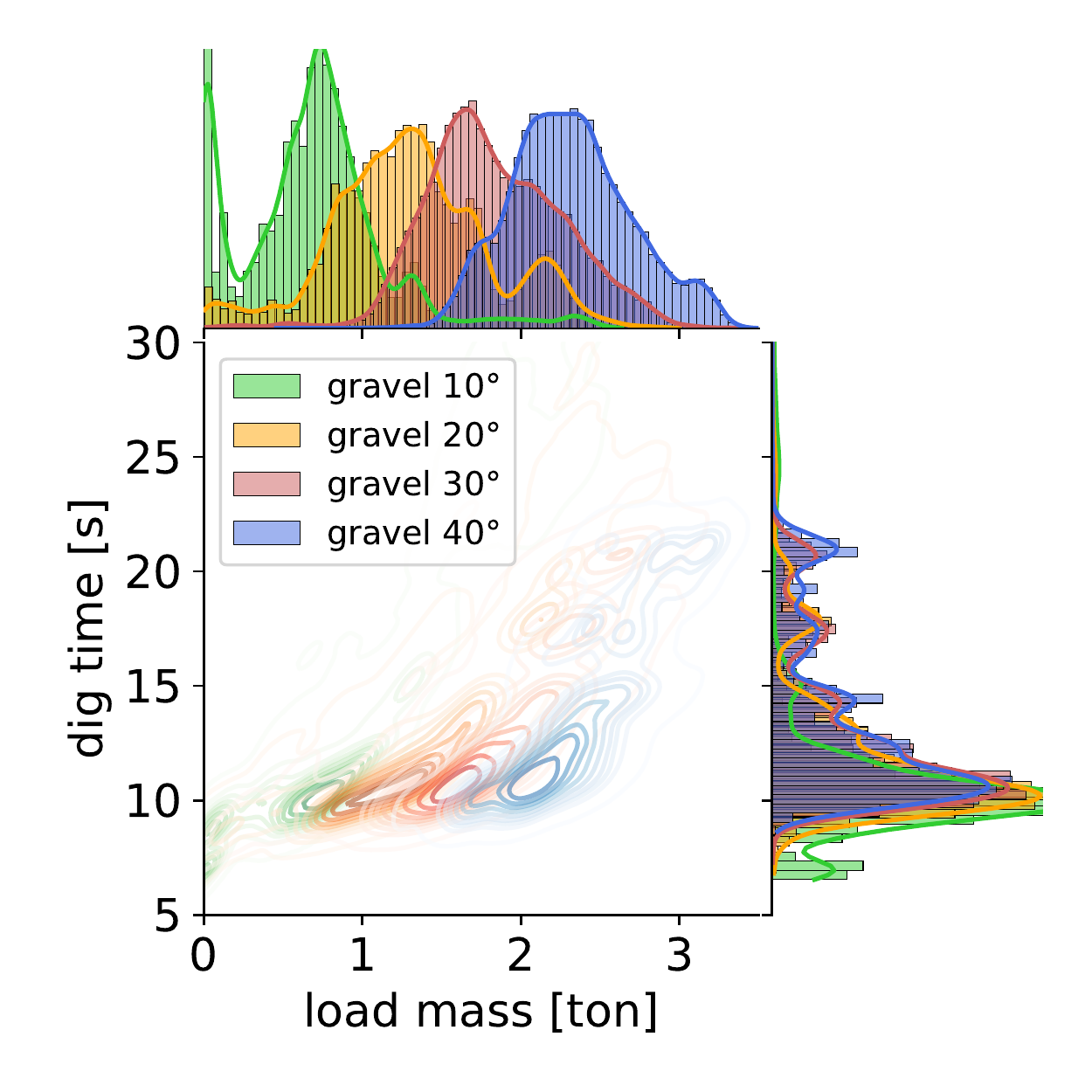} 
            \includegraphics[trim=7 11 15 17, clip, width=40mm]{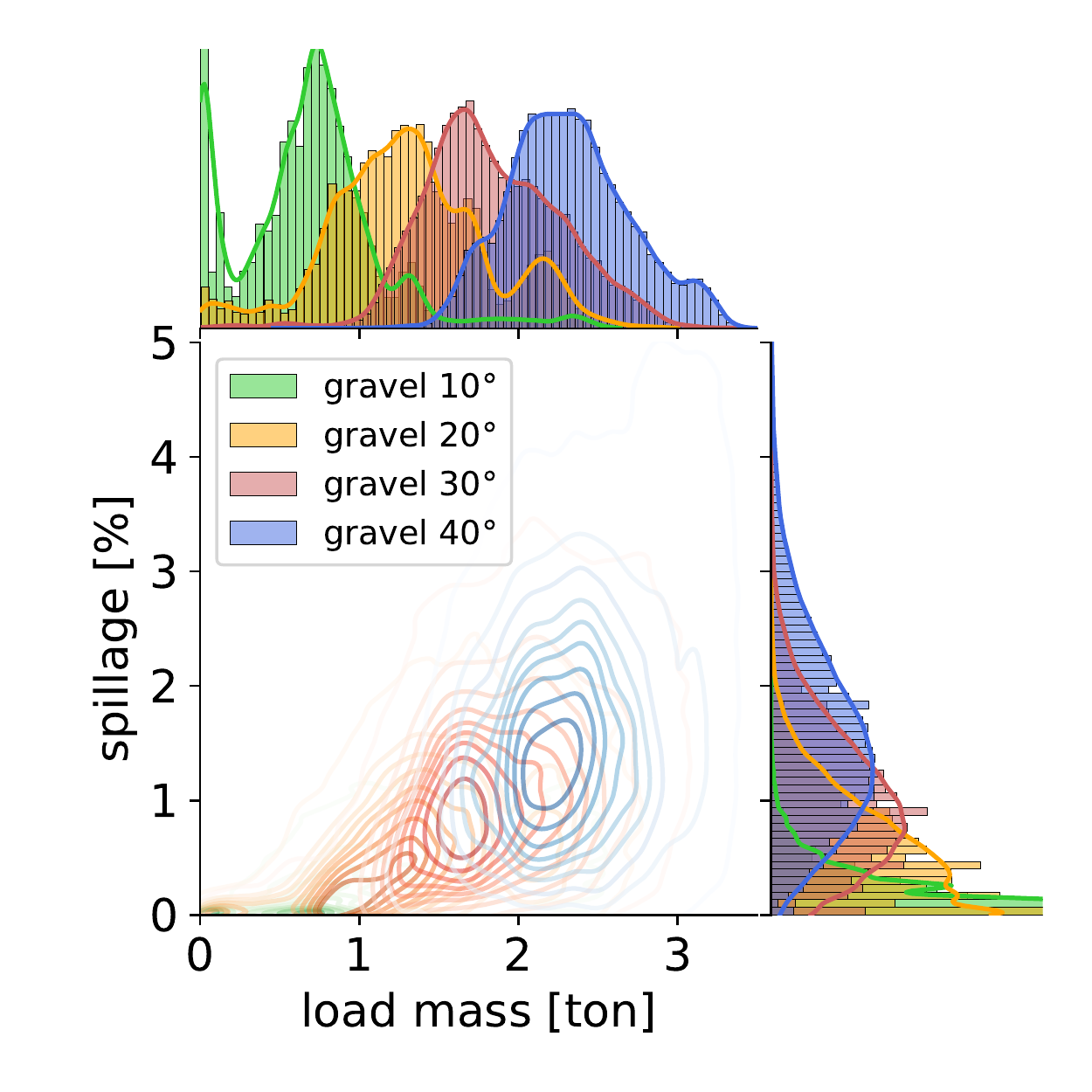}
            \includegraphics[trim=8 11 15 17, clip, width=40mm]{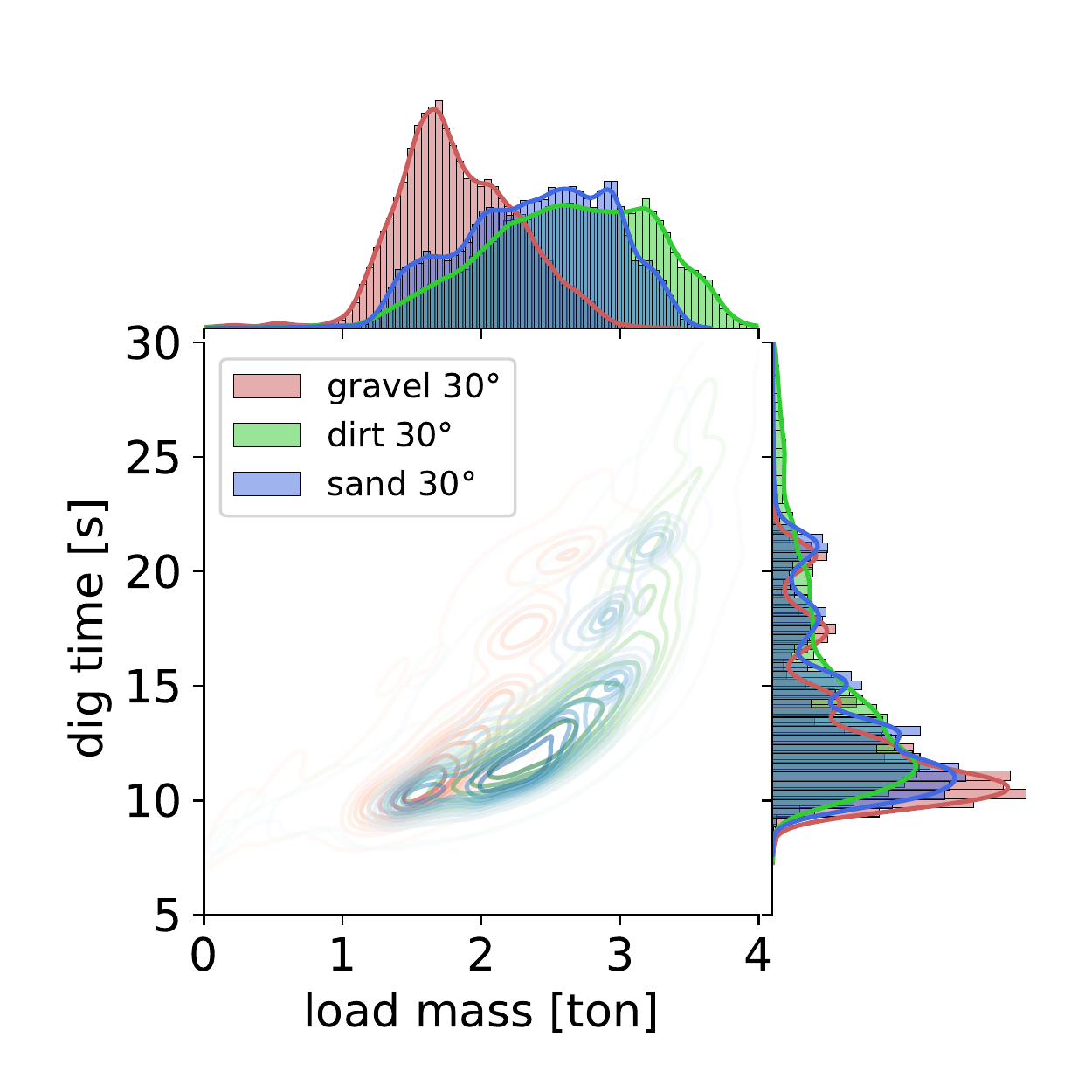}
            \includegraphics[trim=7 11 15 17, clip, width=40mm]{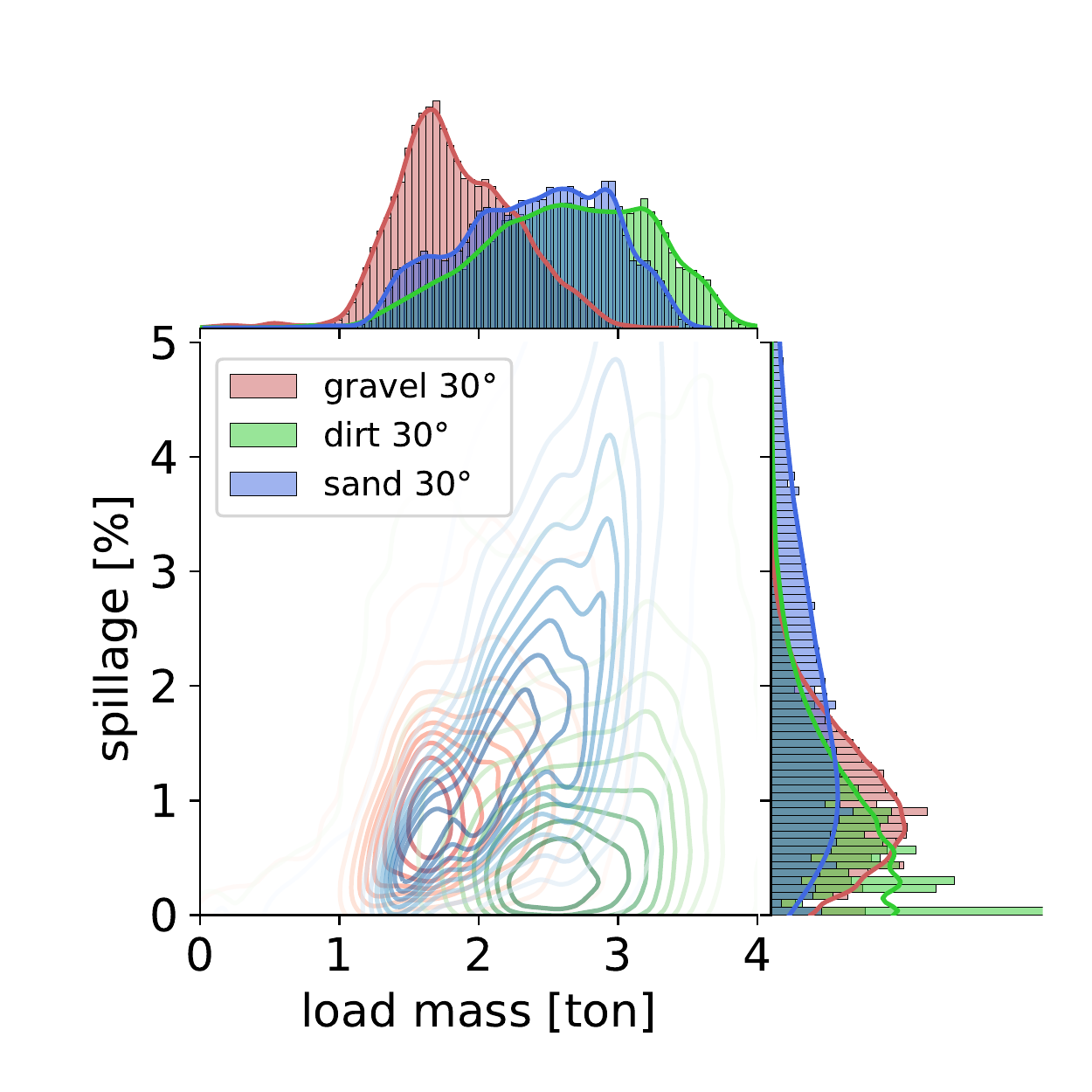}
            \caption{Distribution of loadings relative to mass, time, and spillage for different piles.}
        \label{fig:2d_histogram}
    \end{figure*}

\begin{figure*}
        \centering
        \includegraphics[width=0.9\textwidth, trim = 0 0 0 0, clip]{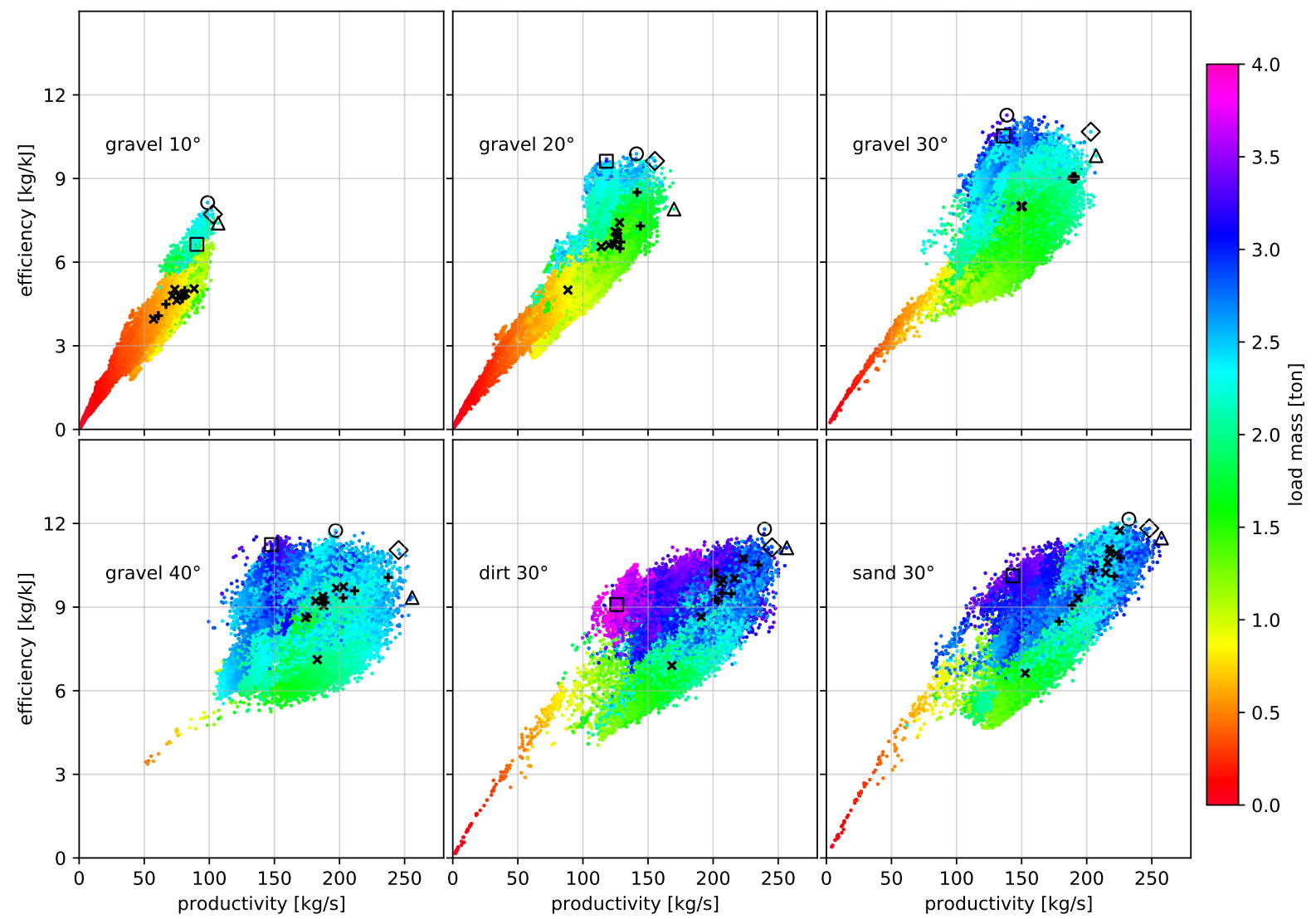} 
        \caption{Distribution of the loading performance for each pile, marking some points of interest; the most efficient ($\circ$);
        productive ($\triangle$); Pareto optimal ($\diamond$); and the highest load mass ($\square$).
        Points highlighted with ($+$) and ($\times$) are loadings performed with two nearly identical sets of action parameters selected
        from gravel $30^\circ$.}
        \label{fig:scatter}
    \end{figure*}

    Figures~\ref{fig:2d_histogram} and \ref{fig:scatter} show the variations in performance over the 270,000 simulated loading operations.
    The distributions over dig time, load mass, 
    and spillage, per pile angle and material, are shown in Figure~\ref{fig:2d_histogram}. 
    In general, the trend is that the load mass increase with the slope, and the dig time is 
    positively correlated with the load mass.  The nominal dig time is around 10-12 s, which can be 
    compared to the 15 s for completing the experimental loading cycle in Figure~\ref{fig:comparison_forces}.
    Higher load mass requires a larger pile slope, but it is harder to achieve a high load with gravel than for 
    sand and dirt. Spillage is mostly below 2\% of the bucket volume.  It increases with the pile slope and is 
    largest for sand and smallest for dirt, presumably thanks to its cohesive property. Figure~\ref{fig:scatter} 
    reveals that, to first order, loading efficiency, productivity, and load mass are linearly related, and they 
    are positively correlated with the pile slope.  However, many of the higher load mass cases are associated 
    with poor productivity.  The efficiency and productivity for dirt and sand have similar distributions, but 
    higher load mass can be achieved with dirt.  The efficiency and productivity are generally lower for gravel 
    than for sand and dirt.

    To study the sensitivity of action parameters, we select two performance points, $(\mathcal{P}_\text{p}, \mathcal{P}_\text{e})$, 
    in the gravel $30^\circ$ data and extract the action parameter values that produce a similar performance.
    The performance of the identical set of action parameters on the five other sets of piles is then highlighted.
    The two performance points in gravel $30^\circ$ are $(150,8.0)$ and $(190,9.0)$, and they are highlighted with 
    ($\times$) and ($+$), respectively.  The corresponding performance points are not gathered narrowly in the distributions 
    for other pile angles but more so for dirt and sand piles with $30^\circ$ slope.  This suggests that loading actions 
    should be adapted to the slope of the pile, while high-performing actions may transfer to piles of different material but similar slope.

  A selection of points of special interest (POI) is made for each pile slope and material. These points correspond 
    to maximum efficiency ($\circ$), productivity ($\triangle$), load mass ($\square$), and a Pareto optimal point
    ($\diamond$). The Pareto optimal point ($\diamond$) is chosen in between maximum efficiency ($\circ$) and productivity ($\triangle$) arbitarily.
    The action parameters and performance values for the POI:s for $30^\circ$ slopes are presented in Table 
    \ref{tab:parameters}. We observe that a high entry speed ($\alpha_1$) and medium-high dig speed ($\alpha_2$) are 
    beneficial for high productivity while high efficiency relates to low dig speed ($\alpha_2$).
    For large load mass, it appears important to trigger tilting at a large digging forces 
    ($\alpha_4$) and to tilt at low speed ($\alpha_6$).
	
    No obvious relations are found for many of the action parameters and performance values.

    \begin{table*}
        \small
        \centering
        \caption{Action parameters and performance for selected loadings on piles with $30^\circ$ slope.}
        \label{tab:parameters}
        \begin{tabular}{l c r r r r r r r r r r r r r} \hline
            soil   & loading     & $\alpha_1$  & $\alpha_2$ & $\alpha_3$    & $\alpha_4$    & $\alpha_5$    & $\alpha_6$    & $\alpha_7$    & $\alpha_8$    & $m_\text{load}$   & $t_\text{load}$ & $s_\text{load}$ & $\mathcal{P}_\text{prod}$ & $\mathcal{P}_\text{eff}$ \\  \hline
            gravel & $\circ$ & $0.6$ & $0.2$ & $0.0$ & $1.2$ & $0.4$ & $0.2$ & $-30^\circ$ & $45^\circ$ & $3.40$ & $24.5$ & $1.5$ & $139$ & $\bm{11.27}$ \\
            gravel & $\triangle$ & $0.8$ & $0.4$ & $0.0$ & $0.9$ & $0.6$ & $1.0$ & $-30^\circ$ & $30^\circ$ & $2.15$ & $10.3$ & $2.0$ & $\bm{207}$ & $9.82$ \\
            gravel & $\diamond$ & $0.6$ & $0.4$ & $0.0$ & $1.2$ & $0.4$ & $0.6$ & $-30^\circ$ & $30^\circ$ & $2.51$ & $12.3$ & $0.3$ & $203$ & $10.68$ \\
            gravel & $\square$ & $0.6$ & $0.4$ & $0.3$ & $1.2$ & $0.8$ & $0.2$ & $-30^\circ$ & $45^\circ$ & $\bm{3.41}$ & $25.1$ & $2.4$ & $136$ & $10.53$ \\
            \hline
            dirt & $\circ$ & $0.6$ & $0.2$ & $0.0$ & $0.9$ & $0.8$ & $1.0$ & $-30^\circ$ & $30^\circ$ & $2.81$ & $11.7$ & $0.8$ & $240$ & $\bm{11.80}$ \\
            dirt & $\triangle$ & $0.8$ & $0.6$ & $0.9$ & $1.2$ & $0.2$ & $1.0$ & $-40^\circ$ & $30^\circ$ & $2.92$ & $11.3$ & $0.0$ & $\bm{257}$ & $11.13$ \\
            dirt & $\diamond$ & $0.8$ & $0.4$ & $0.3$ & $0.9$ & $0.2$ & $0.8$ & $-30^\circ$ & $30^\circ$ & $2.76$ & $11.2$ & $1.3$ & $245$ & $11.14$ \\
            dirt & $\square$ & $0.6$ & $0.4$ & $0.3$ & $1.2$ & $0.6$ & $0.2$ & $-30^\circ$ & $45^\circ$ & $\bm{4.12}$ & $32.7$ & $1.5$ & $126$ & $9.09$ \\
            \hline
            sand & $\circ$ & $0.8$ & $0.2$ & $0.0$ & $0.9$ & $1.0$ & $1.0$ & $-40^\circ$ & $30^\circ$ & $2.43$ & $10.5$ & $1.8$ & $232$ & $\bm{12.16}$ \\
            sand & $\triangle$ & $0.8$ & $0.6$ & $0.6$ & $0.3$ & $0.2$ & $0.8$ & $-40^\circ$ & $45^\circ$ & $2.98$ & $11.6$ & $3.6$ & $\bm{257}$ & $11.48$ \\
            sand & $\diamond$ & $0.8$ & $0.4$ & $0.0$ & $0.6$ & $1.0$ & $1.0$ & $-30^\circ$ & $45^\circ$ & $2.80$ & $11.2$ & $4.9$ & $248$ & $11.83$ \\
            sand & $\square$ & $0.8$ & $0.4$ & $0.0$ & $1.2$ & $0.6$ & $0.2$ & $-30^\circ$ & $45^\circ$ & $\bm{3.65}$ & $25.5$ & $5.5$ & $143$ & $10.12$ \\
            \hline
             &  &  &  &  &  &  &  &  &  & [ton] & [sec] & [\%] &  [kg/s] & [kg/kJ] \\
        \end{tabular}
    \end{table*}

\begin{figure*}
        \raggedright
        \includegraphics[trim=5 25 5 7, clip, width=39mm, height=22mm]{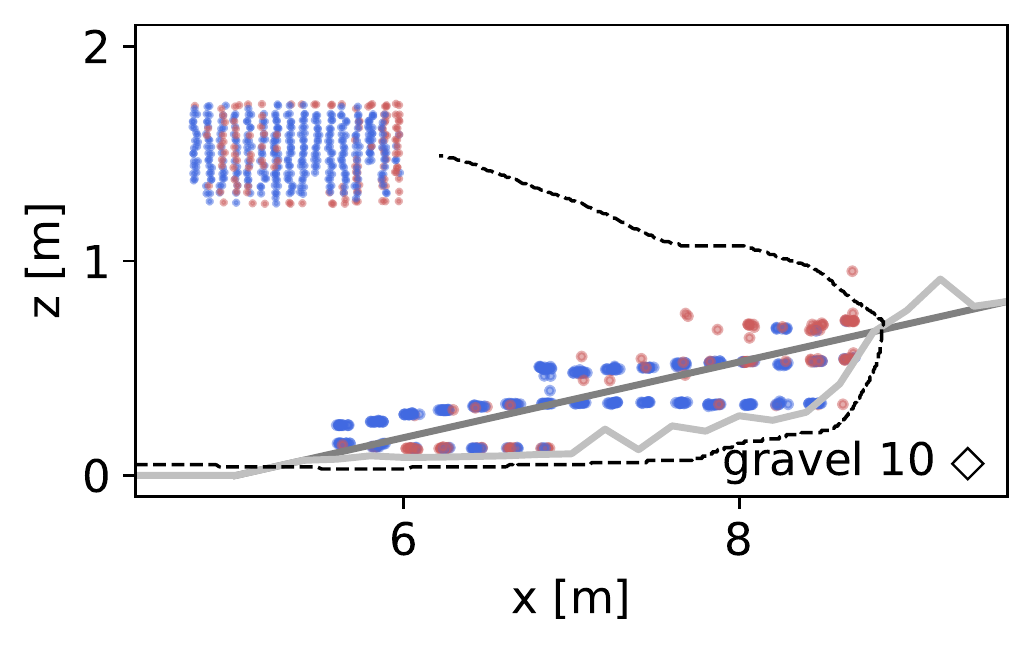}
        \includegraphics[trim=35 25 5 7, clip, width=35mm, height=22mm]{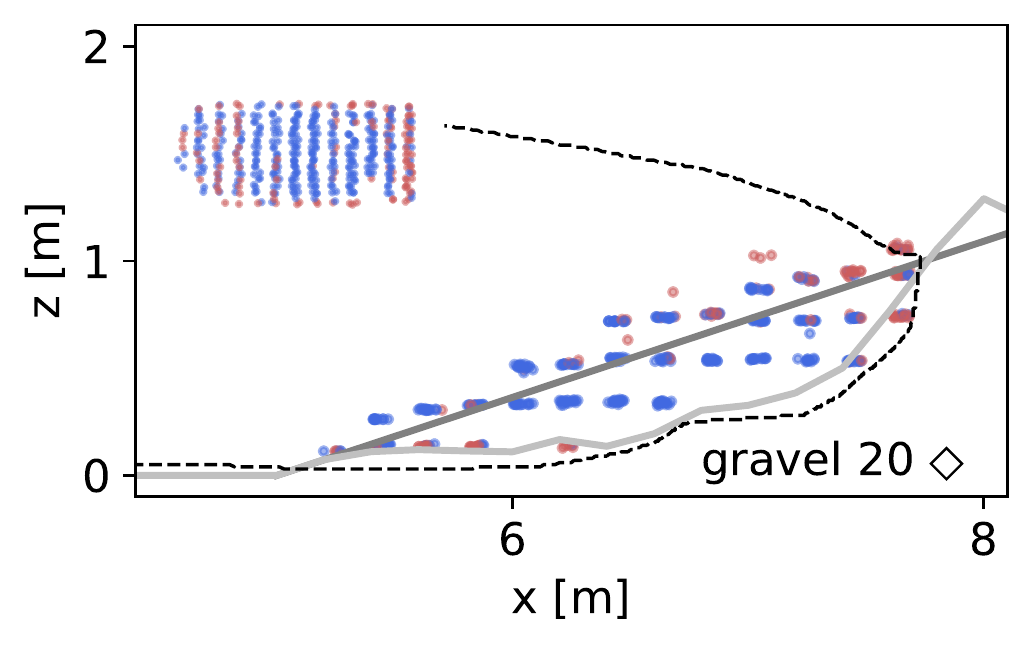} 
        \includegraphics[trim=35 25 5 7, clip, width=35mm, height=22mm]{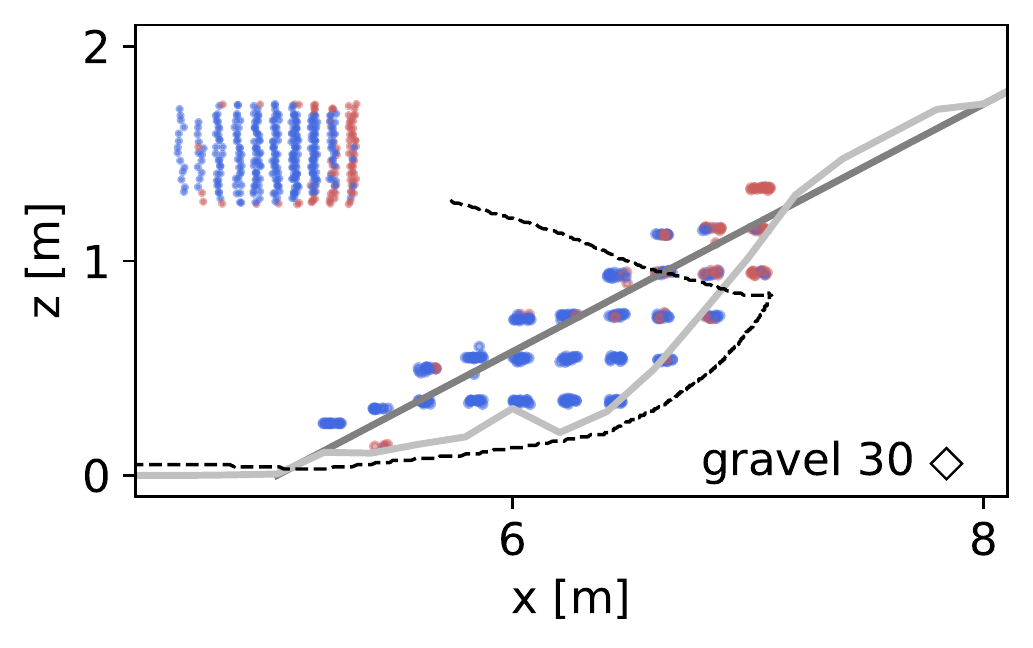}
        \includegraphics[trim=35 25 5 7, clip, width=35mm, height=22mm]{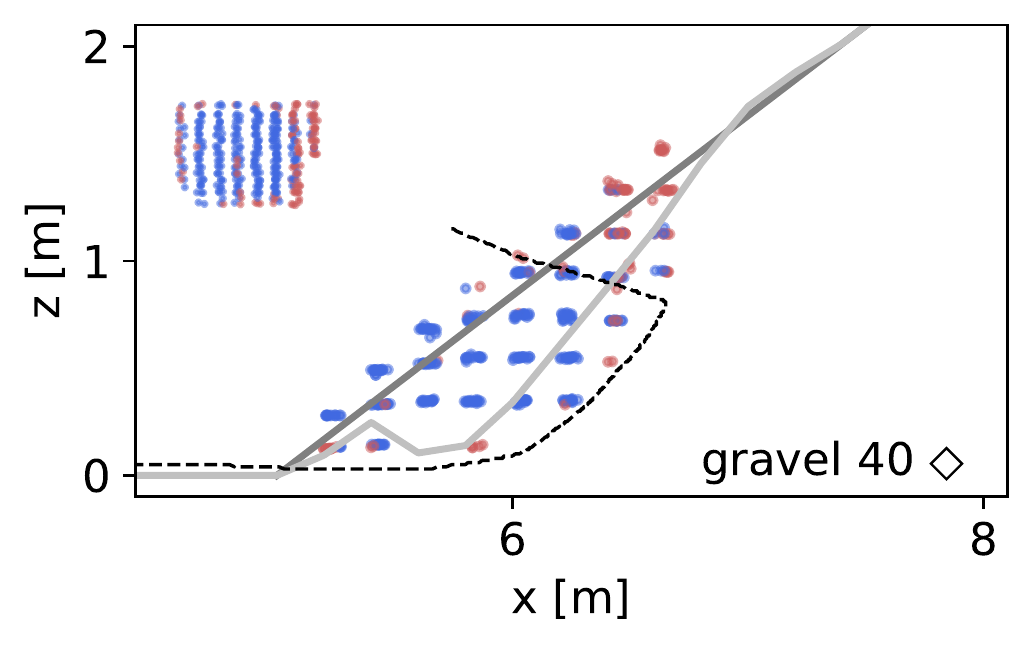} 
        \raisebox{8mm}{\includegraphics[trim=6 14 6 5, clip, height=14mm]{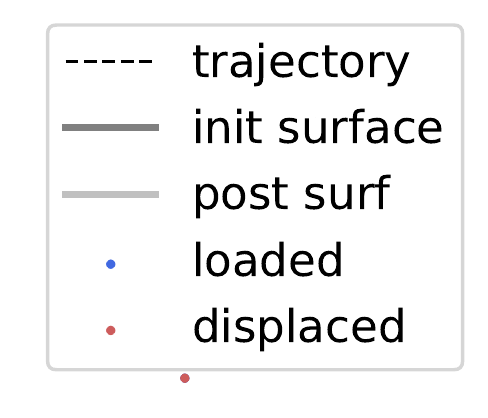}}\\
        \caption{Bucket tip trajectories for Pareto optimal loadings from gravel piles with different slopes. 
        The initial and resulting pile surface is shown as well as what mass is loaded successfully or just displaced,
        with the top view projection included in the upper left corners. }
        \label{fig:particles_slope}
    \end{figure*}

\begin{figure*}
        \raggedright
        \includegraphics[trim=5 40 5 7, clip, width=39mm, height=20mm]{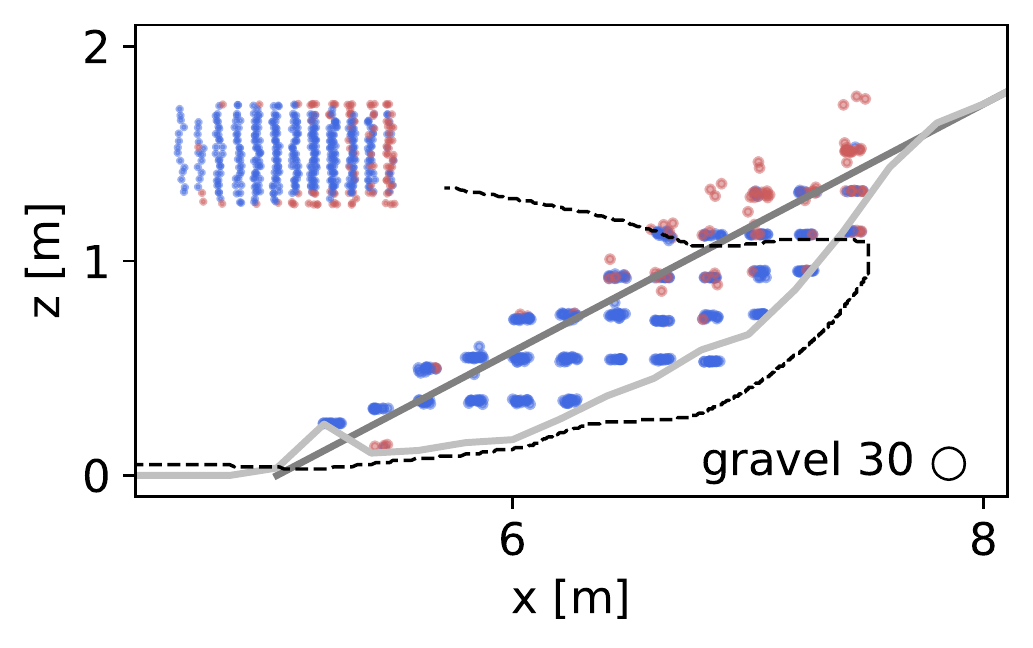}
        \includegraphics[trim=35 40 5 7, clip, width=35mm, height=20mm]{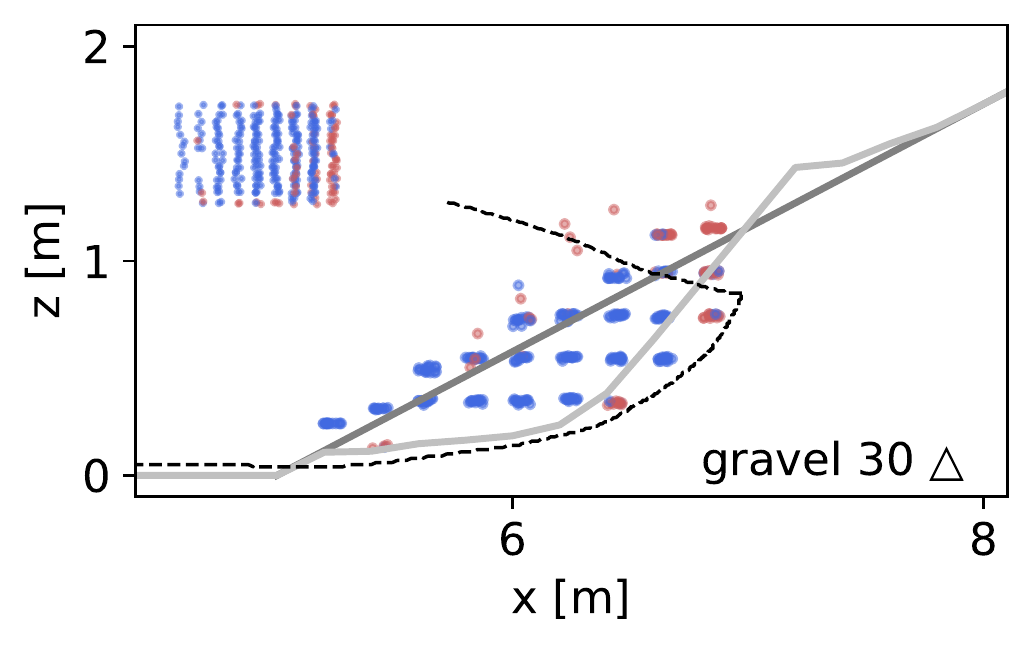}
        \includegraphics[trim=35 40 5 7, clip, width=35mm, height=20mm]{trajectory_3_G30.pdf}
        \includegraphics[trim=35 40 5 7, clip, width=35mm, height=20mm]{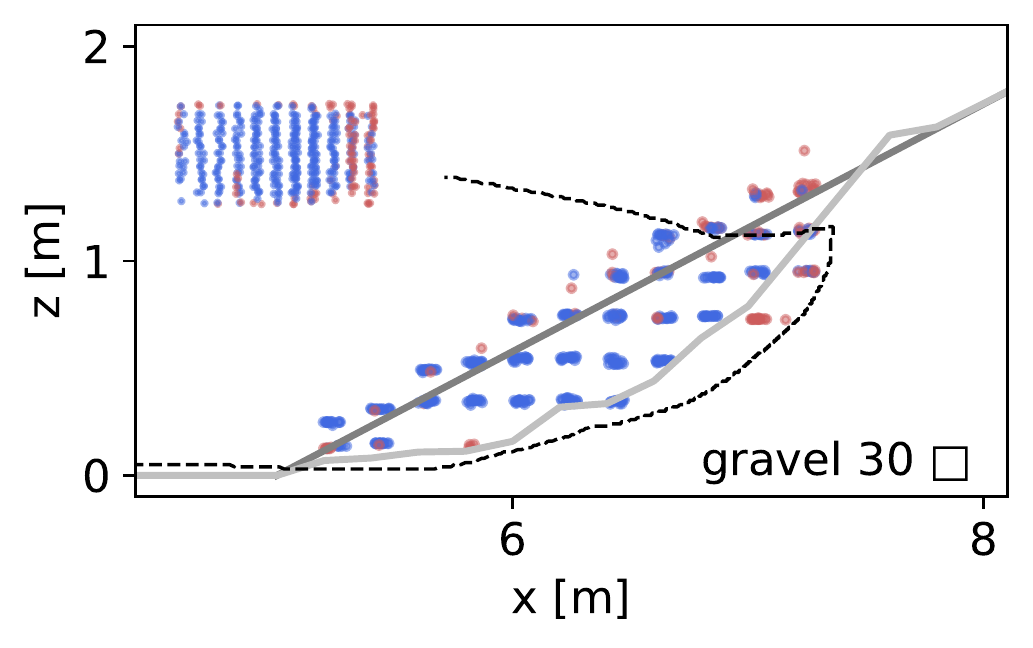}
        \raisebox{8mm}{\includegraphics[trim=6 14 6 5, clip, height=12mm]{trajectory_legend.pdf}}\\
        \includegraphics[trim=5 40 5 7, clip, width=39mm, height=20mm]{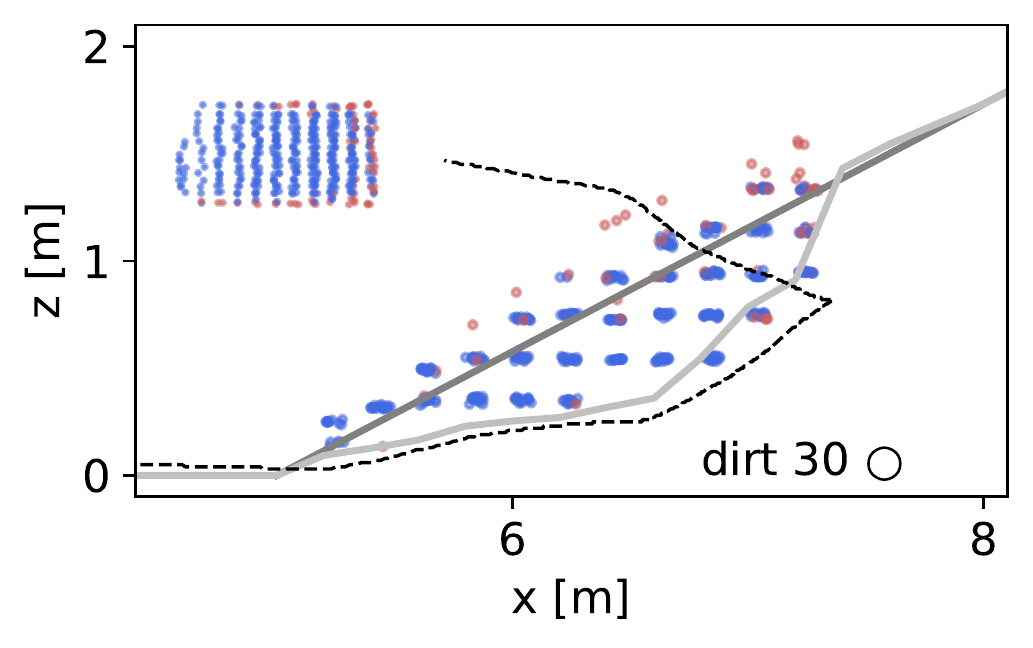}
        \includegraphics[trim=35 40 5 7, clip, width=35mm, height=20mm]{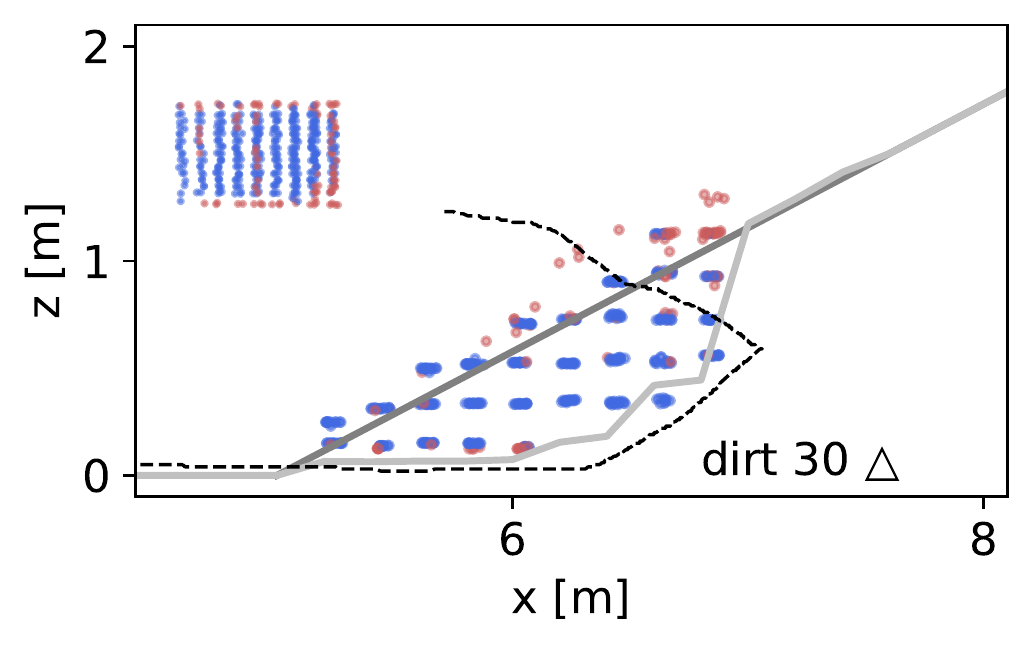}
        \includegraphics[trim=35 40 5 7, clip, width=35mm, height=20mm]{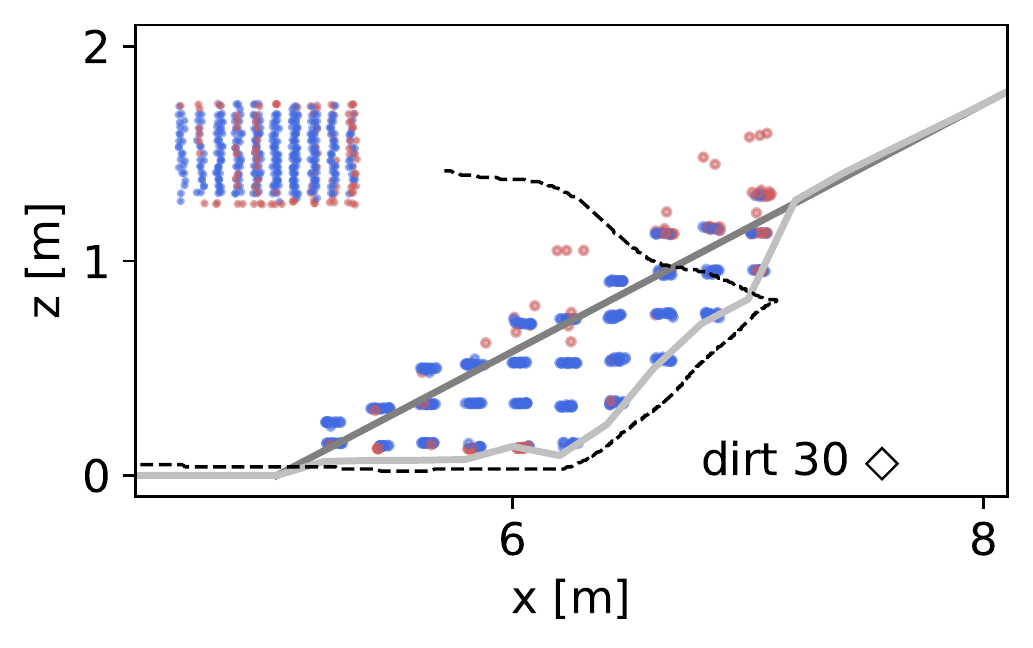}
        \includegraphics[trim=35 40 5 7, clip, width=35mm, height=20mm]{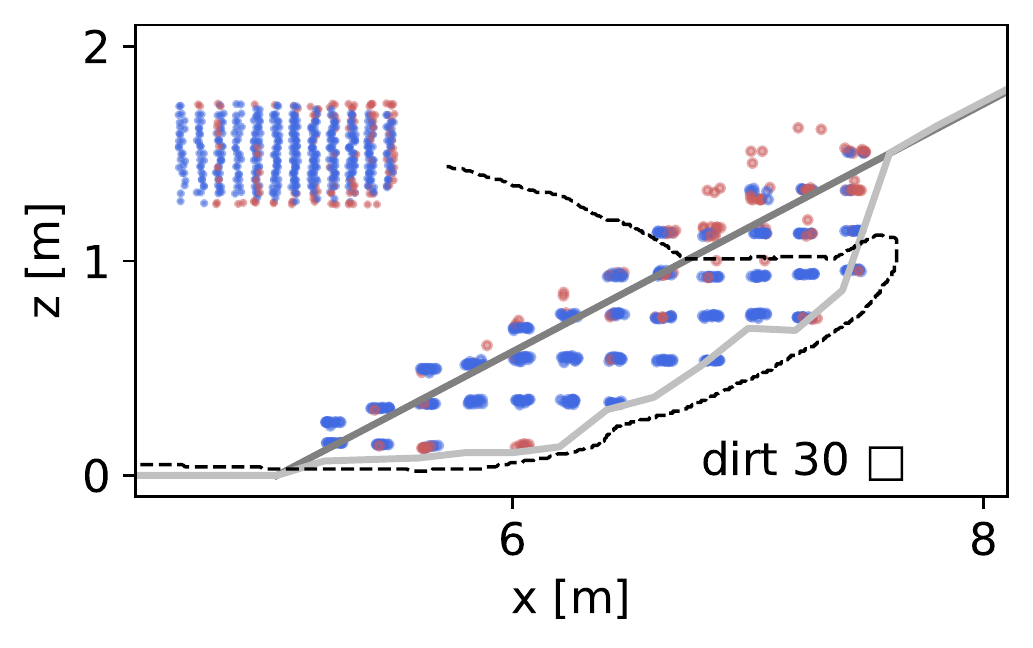}
        \raisebox{8mm}{\includegraphics[trim=6 14 6 5, clip, height=12mm]{trajectory_legend.pdf}}\\
        \includegraphics[trim=5 7 5 3, clip, width=39mm, height=25mm]{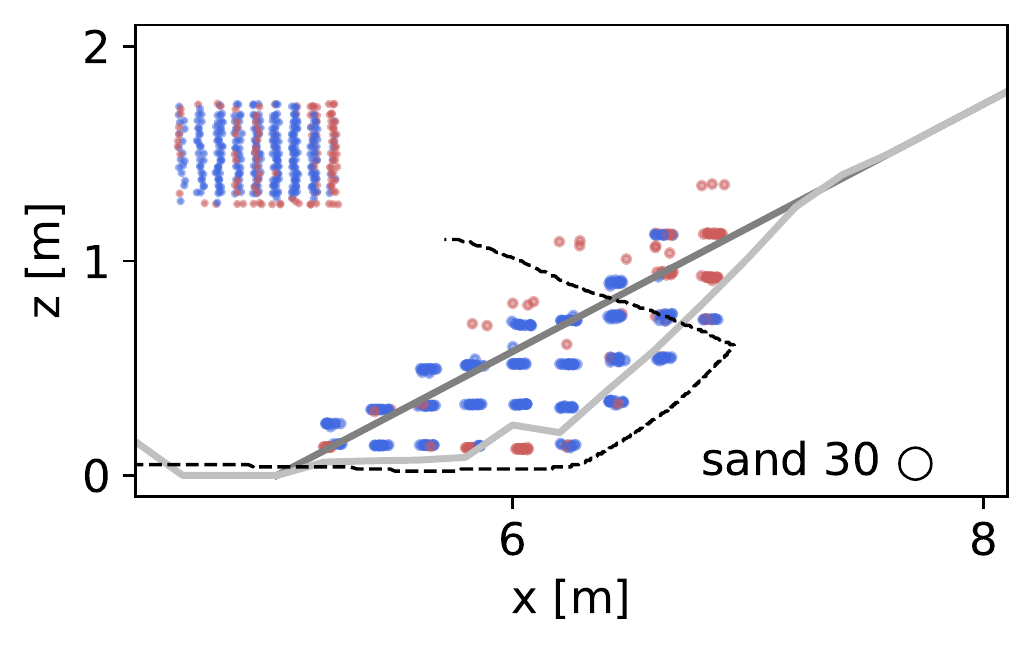}
        \includegraphics[trim=35 7 5 3, clip, width=35mm, height=25mm]{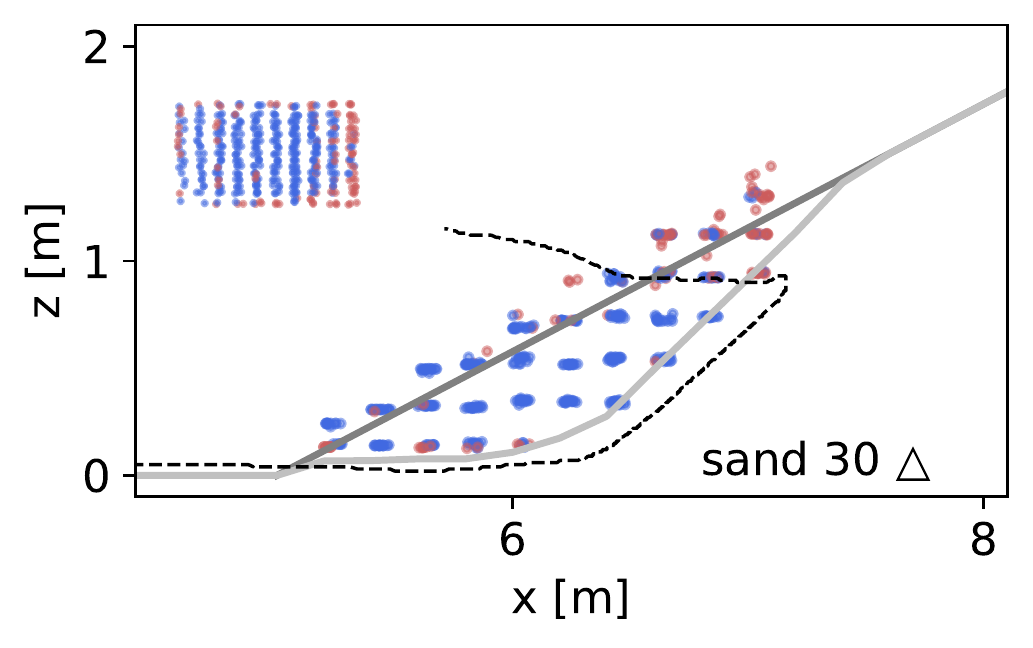}
        \includegraphics[trim=35 7 5 3, clip, width=35mm, height=25mm]{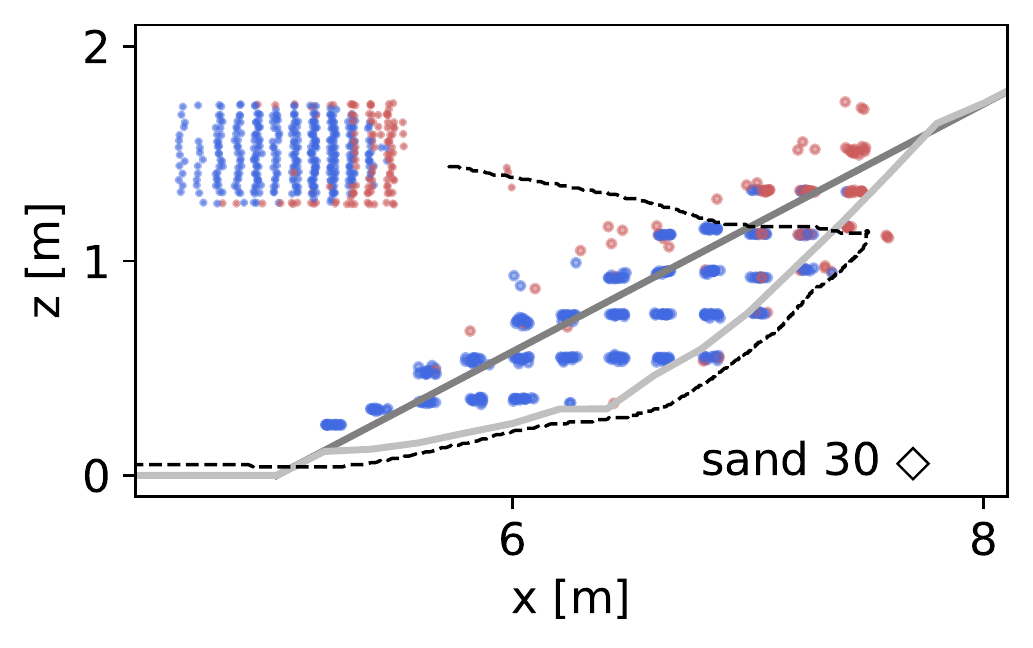}
        \includegraphics[trim=35 7 5 3, clip, width=35mm, height=25mm]{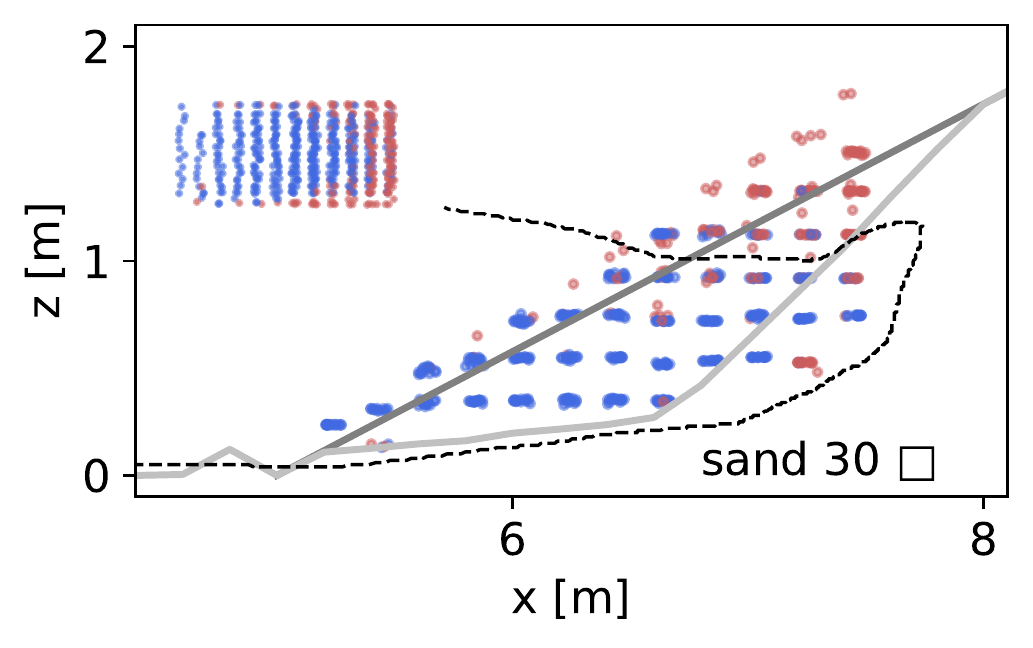}
        \raisebox{12mm}{\includegraphics[trim=6 14 6 5, clip, height=12mm]{trajectory_legend.pdf}}\\
        \caption{Bucket tip trajectories for $30^\circ$ piles of different material (row) and the four different loading performance POI (column).
        The initial and resulting pile surface is shown as well as what mass is loaded successfully or just displaced, 
        with the top view projection included in the upper left corners. }
        \label{fig:particles_material}
    \end{figure*}

    Although it is in general not possible to control the bucket to follow a prescribed path, it is interesting to 
    analyze the trajectories of the POI loadings. This is presented in Figure \ref{fig:particles_slope} and 
    \ref{fig:particles_material} for different slope and material, respectively. Also shown are the initial and 
    resulting pile surfaces as well as the initial position of the mass that is loaded or just displaced by the loading action. 
    At lower slope, in Figure \ref{fig:particles_slope}, we note a larger tendency for soil being pushed forward and 
    not ending up in the bucket, negatively affecting efficiency, productivity, and load mass. 
    In Figure \ref{fig:particles_material}, the trajectories for each type of POI (column) are shown for the different materials (row).
    The loadings of maximal mass ($\square$) are characterized by digging deeper into the pile.
    For the other POI loadings, there are no obvious trends when comparing only the trajectories.


    \section{Discussion}
    \label{sec:discussion}
    Overall, the admittance-like control method seems to work well if adjusted for the pile slope.
    The results suggest that the preferred digging actions should preserve and exploit a steep pile slope.
    It appears more important to adapt the loading actions to the pile shape than to 
    the soil type, at least among the materials tested in this study. High digging speed favors 
    high productivity, while energy-efficient loading requires a lower dig speed.

    Several delimitations have been made in the present study, and many questions are left for future 
    work. The effect of more complex pile shapes and other materials needs to be studied. The reason 
    for the moderate load mass for gravel is not understood. Possibly the virtual gravel represents 
    a more densely packed and stronger material than what was used in the field experiments. 
    Also, from field tests, one expects a larger difference in the high-performance trajectories 
    between the materials, e.g., longer, and more shallow digging for dirt than gravel and lower and 
    more deep thrusting motions for sand. These tendencies are present for the high-productivity 
    trajectories in Figure \ref{fig:particles_material}, but the difference is expected to be larger. Digging actions can be 
    represented and discretized differently than in the present study, and it is certainly possible 
    that higher performance loading can be discovered within the present action space. The spillage and 
    resulting pile surface can be observed in the results, but we have not investigated how this 
    penalizes sequential loadings. 
    
\section{Conclusion}
\label{sec:conclusion}
    We have develope a simulator to explore the sequential loading actions which maximize the performance of automated 
    wheel loader systems.
    The simulator is based on 3D multibody dynamics and deformable terrain with real-time capability.
    A vast number of loading simulations demonstrates that the combined action and pile state significantly affects the performance. 
    As the next step, we will study the sequential loading scenario and address the optimization problem.

\section*{Acknowledgements}
This work has in part been supported by Komatsu Ltd and Algoryx Simulation AB.  The simulations were 
performed on resources provided by the Swedish National Infrastructure for Computing 
(SNIC dnr 2021/5-234) at High Performance Computing Center North (HPC2N).
    
\bibliography{ISARC_arxiv}

\end{document}